\documentclass[twocolumn]{aastex701}
\usepackage{CJK}
\usepackage{siunitx}

\usepackage{graphicx}
\usepackage{subcaption}

\begin{document}
\begin{CJK*}{UTF8}{gbsn}

\title{Post-perihelion Coma Composition of the Interstellar Comet 3I/ATLAS from Optical Spectroscopy}

\correspondingauthor{Ruining Zhao and Xiliang Zhang}

\author[0000-0003-4936-4959]{Ruining Zhao (赵瑞宁)}
\affiliation{CAS Key Laboratory of Optical Astronomy, National Astronomical Observatories, Chinese Academy of Sciences, Beijing 100101, China}
\email{rnzhao@nao.cas.cn}

\author{Xiliang Zhang (张西亮)} 
\affiliation{Yunnan Observatories, Chinese Academy of Sciences (CAS), Kunming 650216, China}
\affiliation{School of Astronomy and Space Sciences, University of Chinese Academy of Sciences, Beijing 100049, China}
\email{Zhangxiliang@ynao.ac.cn}

\author[0000-0002-5033-9593]{Bin Yang (杨彬)}
\affiliation{Instituto de Estudios Astrof\'isicos, Facultad de Ingenier\'ia y Ciencias, Universidad Diego Portales, Santiago, Chile}
\affiliation{Planetary Science Institute, 1700 E Fort Lowell Rd STE 106, Tucson, AZ 85719, USA}
\email{}

\author{Xiangyu Fan} 
\affiliation{Yunnan Observatories, Chinese Academy of Sciences (CAS), Kunming 650216, China}
\affiliation{School of Astronomy and Space Sciences, University of Chinese Academy of Sciences, Beijing 100049, China}
\email{}

\author[0000-0003-4489-9794]{Shu Wang (王舒)}
\affiliation{CAS Key Laboratory of Optical Astronomy, National Astronomical Observatories, Chinese Academy of Sciences, Beijing 100101, China}
\email{shuwang@nao.cas.cn}

\author{Yang Huang (黄样)}
\affiliation{School of Astronomy and Space Sciences, University of Chinese Academy of Sciences, Beijing 100049, China}
\email{}

\author{Jifeng Liu (刘继峰)}
\affiliation{CAS Key Laboratory of Optical Astronomy, National Astronomical Observatories, Chinese Academy of Sciences, Beijing 100101, China}
\affiliation{School of Astronomy and Space Sciences, University of Chinese Academy of Sciences, Beijing 100049, China}
\affiliation{Institute for Frontiers in Astronomy and Astrophysics, Beijing Normal University, Beijing 102206, China}
\affiliation{New Cornerstone Science Laboratory, National Astronomical Observatories, Chinese Academy of Sciences, Beijing 100012, China}
\email{}

\begin{abstract}
We present multi-epoch optical spectroscopy of the interstellar comet 3I/ATLAS obtained between December 2025 and January 2026 (heliocentric distances 1.8--3.3 au), yielding post-perihelion production rates and mixing ratios for CN, C$_3$, C$_2$, CH, and gaseous metals (Fe\,{\sc i} and Ni\,{\sc i}). Our results show that the coma is less depleted in C$_2$ after perihelion than before, indicative of subsurface activation or compositional heterogeneity. The outgassing profiles reveal a pronounced perihelion asymmetry: CN and metal production rates decline more gradually outbound than inbound, consistent with the reported behavior of H$_2$O and implying a change in the comet's activity pattern across perihelion. Despite being metal-rich relative to its H$_2$O content, 3I follows the metal--CO correlation observed in comets of diverse origins, suggesting that gaseous metal release is more closely linked to a CO-bearing volatile reservoir than to H$_2$O, potentially in the form of metal carbonyls. In addition, the [O\,{\sc i}] $\lambda6300$ emission shows a significant residual after subtracting the expected contributions from H$_2$O, CO$_2$, and CO, which may reflect systematic uncertainties in the photodissociation yields of those molecules or a contribution from additional oxygen-bearing parents.

\end{abstract}



\section{Introduction} \label{sec:intro}

% Comets formed in the protoplanetary disk and retain a record of the physical and chemical conditions present during the earliest stages of planet formation. Large surveys of solar system comets have shown that, despite measurable compositional differences, their composition remains braodly similar, with volatile abundances typically vary within about an order of magnitude \citep{Mumma:2011}, suggesting generally a common origin. 

Interstellar objects （ISOs) are small bodies that formed in the protoplanetary disks of other stars and were later ejected into interstellar space \citep{Jewitt:2023,Fitzsimmons:2024}. Their rare visits to the solar system provide a unique opportunity to sample the physical and chemical properties of distant planetary systems. Comparing these objects with solar system comets places our own system in a broader context and highlights the key factors that govern planet formation across diverse environments.

% The first known ISO, 1I/`Oumuamua, was discovered in 2017 \citep{2017CBET.4450....1W,Meech:2017}. Although it exhibited a measurable non-gravitational acceleration likely caused by comet-like outgassing \citep{Micheli:2018}, no gas species were detected spectroscopically, leaving its composition largely unconstrained \citep{Ye:2017,Fitzsimmons:2018}. 
% By contrast, the second ISO, 2I/Borisov, was discovered in 2019 with a well-developed coma that allowed detailed compositional studies \citep{Borisov:2019}.
% It was found to have a very high CO abundance relative to water compared with most solar system comets, placing it among the most CO-rich comets known \citep{Bodewits:2020,Cordiner:2020}. 
% Moreover, its high linear polarization resembled that of C/1995 O1 (Hale-Bopp), suggesting a relatively pristine nature \citep{Bagnulo:2021}. These properties underscore the physical and chemical distinctiveness of 2I/Borisov and highlight the potential of ISOs to probe the diversity of planetary formation environments.

The third ISO, 3I/ATLAS (hereafter 3I), was discovered in July 2025 and passed perihelion in late October of the same year \citep{Denneau:2025}. 
During the inbound phase, observations with the James Webb Space Telescope (JWST) revealed a CO$_2$-dominated coma at 3.32\,au, prior to the onset of significant H$_2$O sublimation \citep{Cordiner:2025}. Early optical spectroscopic observations indicated a strongly carbon-chain depleted composition based on upper limits on C$_2$ \citep{SalazarManzano:2025,2026ApJ...998L..30L}. Subsequent detections of C$_2$ indicated a less extreme depletion, with most reported $\log(Q_{\rm C_2}/Q_{\rm CN})$ values clustering around $-0.4$ \citep{Hutsemekers:2026}.
Atomic nickel (Ni\,{\sc i}) was first detected at a heliocentric distance of $r_{\rm h}=3.78$\,au, whereas atomic iron (Fe\,{\sc i}) was identified only later at $r_{\rm h}=2.64$\,au, yielding an initially elevated Ni/Fe ratio that later approached the typical values observed in solar system comets \citep{2025ApJ...995L..34R, 2025arXiv251011779H, 2025arXiv251209020H, Hutsemekers:2026}. 

Following perihelion, the carbon-chain depletion diminished further, as indicated by the reported range of $\log(Q_{\rm C_2}/Q_{\rm CN}) = -0.26$--0.05 \citep{Jehin:2025,Jehin:2025b,2026arXiv260116983H,Kawakita:2026}. Infrared spectroscopy revealed strong CO emission, suggesting a transition from CO$_2$- to CO-dominated supervolatile outgassing  \citep{Roth:2026b,Cordiner:2026}. The abundances of several organic molecules, specifically CH$_3$OH and CH$_4$, were also found to be enhanced relative to solar-system comets, possibly reflecting a carbon- and oxygen-rich environment \citep{Roth:2026b,Belyakov:2026,Cordiner:2026}. 
Isotopic studies revealed strongly enriched D/H along with elevated carbon and nitrogen isotopic ratios, indicating formation in a very cold, distant, and relatively metal-poor environment, possibly within an ancient planetary system. \citep{SalazarManzano:2026,Cordiner:2026,Opitom:2026,Roth:2026}.

Taken together, these studies established an unusual and evolving picture of the coma composition of 3I. However, most of them were effectively snapshots obtained at isolated epochs, making it difficult to understand how the coma of 3I changed in detail over time. Multi-epoch observations are therefore essential. To directly track its post-perihelion evolution, we present multi-epoch optical spectroscopic observations of 3I obtained with 2-m class telescopes after perihelion, allowing us to directly trace the temporal evolution of the coma composition during the outbound phase. The observation and data reduction are described in Section \ref{sec:obs}. The reduced spectra are modeled and analyzed in Section \ref{sec:analysis}. The results are discussed in Section \ref{sec:dis} and summarized in Section \ref{sec:sum}.

\section{Observation and Data Reduction} \label{sec:obs}

\subsection{2.16-meter telescope}

We obtained four long-slit spectra of 3I in 2025 December with the Beijing Faint Object Spectrograph and Camera (BFOSC) mounted on the Xinglong 2.16-meter Telescope. The instrument is equipped with a $2{\rm K}\times2{\rm K}$ CCD featuring a pixel scale of $0\farcs274$ \citep{2016PASP..128k5005F}. A $2\farcs3\times9\farcm4$ slit was used and fixed in the north-south direction. As the observations were conducted within $\sim$1\,hr of transit, this orientation remained nearly aligned with the parallactic angle, thereby minimizing atmospheric dispersion. The G3 grism, covering 330--660\,nm, was used, delivering a resolving power of $R\sim320$ when combined with the $2\farcs3$ slit. Bias frames, dome flats, and Fe-Ar arc lamp exposures were obtained each afternoon. During the night, 3I was observed together with a nearby solar analog and a flux standard for calibration. The solar analogs were selected from the LAMOST solar-type star catalog \citep{2020ApJ...894L..11Z} and have been verified to exhibit solar-like spectra. Slit-view images were obtained for each target prior to science exposures and were used in the reduction to estimate and correct for slit losses caused by seeing variations. A brief observing log is provided in Table~\ref{tab:log}, with the full version given in Table~\ref{tab:full}.

\subsection{2.40-meter telescope}

In addition to the BFOSC observations, we obtained eight long-slit spectra of 3I in 2025 December and 2026 January using the Yunnan Faint Object Spectrograph and Camera (YFOSC) mounted on the 2.4-meter Lijiang telescope. The instrument is equipped with a $2{\rm K}\times4{\rm K}$ CCD featuring a pixel scale of $0\farcs283$ \citep{Wang:2019}. Observations were carried out with the G3 grism and a $2\farcs5 \times 9\farcm4$ slit fixed in the north-south direction. This configuration yields a resolving power of $R\sim280$ over the wavelength range 320--930\,nm. Dome flats were obtained either before or after each observing night. During each night, 3I was observed together with a nearby solar analog and a flux standard. Wavelength calibration was performed using a He-Ne arc lamp and further refined using observations of the planetary nebula NGC~2392. The observing log is also provided in Table~\ref{tab:log}, with the full version given in Table~\ref{tab:full}.

\subsection{Data reduction}

Both the BFOSC and YFOSC data were reduced with the {\sc astro-plpy}\footnote{\url{https://github.com/RuiningZHAO/plpy}\label{fn:plpy}} package \citep{Zhao:2026} following standard procedures for long-slit spectroscopy, with additional treatments described in Appendix~\ref{app:data}. The sky background was estimated from source-free regions far from the comet trace and subtracted. To ensure consistency between the two instruments in the subsequent production rate calculation, we adopted the same on-sky extraction aperture of $22\arcsec$ for both data sets, corresponding to 80 pixels for BFOSC and 78 pixels for YFOSC.

\startlongtable
\begin{deluxetable*}{lccccccrr}
    \tablecaption{Observation log of 3I/ATLAS.\label{tab:log}}
    \tablewidth{0pt}
    \tablehead{
        \colhead{Date of 2025} & \colhead{$\Delta$\tablenotemark{a}} & \colhead{$r_h$\tablenotemark{b}} & \colhead{$\dot{r}_h$\tablenotemark{c}} & \colhead{Tel./Inst.} & \colhead{$\lambda$\tablenotemark{d}} & \colhead{Slit Width} & \colhead{Exposure} & \colhead{Airmass}\\
        \colhead{(UT)} & \colhead{(au)} & \colhead{(au)} & \colhead{(km\,s$^{-1}$)} & \colhead{} & \colhead{} & \colhead{($\arcsec$)} & \colhead{(s)} & \colhead{}
    }
    \startdata
    Dec. 2  & 1.892 & 1.846 & 42.44 & 2.16-m/BFOSC & 330--660 & 2.3      & 4$\times$300 & 1.72--1.61\\
    Dec. 20 & 1.799 & 2.333 & 50.43 & $\cdots$     & $\cdots$ & $\cdots$ & 3$\times$300 & 1.22--1.21\\
    Dec. 23 & 1.807 & 2.423 & 51.28 & $\cdots$     & $\cdots$ & $\cdots$ & 8$\times$300 & 1.18--1.22\\
    Dec. 26 & 1.823 & 2.509 & 52.00 & $\cdots$     & $\cdots$ & $\cdots$ & 8$\times$300 & 1.19--1.26\\
    \hline
    Dec. 4  & 1.872 & 1.898 & 43.71 & 2.40-m/YFOSC & 320--930 & 2.5      & 300          & 1.21\\
    Dec. 6  & 1.855 & 1.947 & 44.78 & $\cdots$     & $\cdots$ & $\cdots$ & 4$\times$300 & 1.78--1.60\\
    Dec. 7  & 1.846 & 1.973 & 45.32 & $\cdots$     & $\cdots$ & $\cdots$ & 3$\times$300 & 1.47--1.40\\
    Dec. 17 & 1.798 & 2.246 & 49.48 & $\cdots$     & $\cdots$ & $\cdots$ & 2$\times$300 & 1.45--1.41\\
    Dec. 19 & 1.798 & 2.306 & 50.15 & $\cdots$     & $\cdots$ & $\cdots$ & 2$\times$300 & 1.07\\
    Dec. 28 & 1.840 & 2.570 & 52.45 & $\cdots$     & $\cdots$ & $\cdots$ & 300          & 1.13\\
    Jan. 11\tablenotemark{e} & 2.064 & 3.002 & 54.76 & $\cdots$     & $\cdots$ & $\cdots$ & 2$\times$300 & 1.13--1.11\\
    Jan. 20\tablenotemark{e} & 2.311 & 3.293 & 55.75 & $\cdots$     & $\cdots$ & $\cdots$ & 600          & 1.27\\
    \enddata
    \tablenotetext{a}{Geocentric distance.}
    \tablenotetext{b}{Heliocentric distance.}
    \tablenotetext{c}{Heliocentric radial velocity.}
    \tablenotetext{d}{Wavelength coverage.}
    \tablenotetext{e}{Date of 2026.}
\end{deluxetable*}

\section{Results and Analysis} \label{sec:analysis}

\begin{figure*}[ht!]
    \centering
    \includegraphics[width=\textwidth]{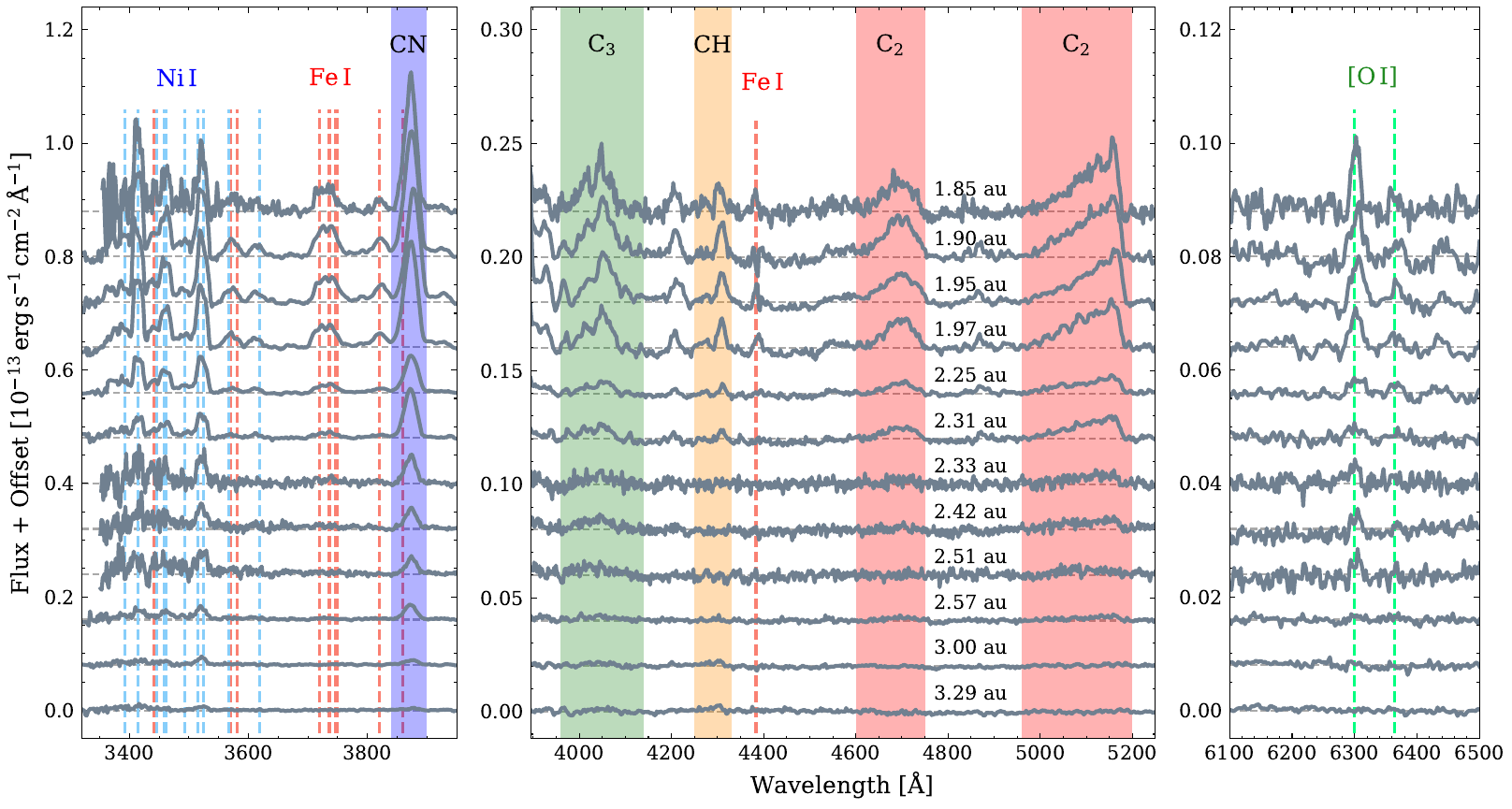}
    \caption{Eimission spectra of 3I/ATLAS taken with BFOSC and YFOSC. For clarity, the full spectral range is divided into three panels with different vertical offsets to emphasize specific features. The left panel highlights emission lines of Ni\,{\sc i} (blue dashed lines) and Fe\,{\sc i} (red dashed lines), and the CN violet band (blue shaded). The middle panel shows molecular bands of C$_3$ (green shaded), CH (orange shaded), and C$_2$ (red shaded), and lists the heliocentric distance for each spectrum. The right panel displays the [O\,{\sc i}] $\lambda\lambda6300,6364$ doublet (green dashed lines). \label{fig:spec}}
\end{figure*}

The optical spectra of 3I exhibit gas emissions superimposed on a dust-scattered solar continuum. To isolate the gas component, we model the continuum by multiplying the spectrum of a solar analog by a low-order polynomial or a B-spline function and then subtract the best-fitting models from the observed spectra. As shown in Figure~\ref{fig:spec}, the resulting pure emission spectra display rich gas emission features, including molecular bands of CN, C$_3$, C$_2$, and CH, atomic lines of Fe\,{\sc i} and Ni\,{\sc i}, and the [O\,{\sc i}] $\lambda\lambda6300,6364$ doublet. In the following subsections, we derive the production rates and mixing ratios for the species related to these emission features and analyze their evolution compared to pre-perihelion measurements.

\subsection{CN, C$_3$, C$_2$, and CH\label{subsec:carbon}}

We first consider the carbon-bearing daughter molecules identified in the spectra. For each species, the band-integrated intensity is measured from the emission spectrum over a fixed wavelength interval: 3840--3900\,${\rm \AA}$ for CN ($\Delta\nu=0$), 3960--4140\,${\rm \AA}$ for C$_3$ ($\Delta\nu=0$), 4960--5200\,${\rm \AA}$ for C$_2$ ($\Delta\nu=0$), and 4250--4330\,${\rm \AA}$ for CH ($\Delta\nu=0$). To avoid contamination from overlapping metal emission, especially near the CN band, the intensities are measured after subtracting the best-fit synthetic metal spectrum described in \S\ref{subsec:metal}. These intensities are then converted to production rates using the Haser model \citep{Haser:1957}, assuming an expansion velocity of $v_{\rm exp}=0.85\,{\rm km\,s^{-1}}\cdot(r_{\rm h}/{\rm au})^{-0.5}$ and adopting the fluorescence efficiencies ($g$-factors) and scale lengths from \citet{Cochran:2012}\footnote{See Tables~\ref{tab:flux} and \ref{tab:para} for summaries of the measured band-integrated intensities and the parameters for Haser modeling, respectively.}. The resulting production rates and mixing ratios relative to CN ($Q_{\rm X}/Q_{\rm CN}$) are listed in Table~\ref{tab:prodrate} on a logarithmic scale. Several measurements beyond $r_{\rm h}\sim2.5$\,au are unavailable because of low signal-to-noise ratios (SNRs).

% Our analysis begins with the carbon-bearing daughter molecules detected in the spectra. 
% Production rates of these daughter molecules are derived using Haser model \citep{Haser:1957}, with an expansion velocity of $v_{\rm exp}=0.85\,{\rm km\,s^{-1}}\cdot(r_{\rm h}/{\rm au})^{-0.5}$, and the $g$-factors and scale lengths from \citet{Cochran:2012}. The production rates of CN ($Q_{\rm CN}$) and the mixing ratios of other species relative to CN ($Q_{\rm X}/Q_{\rm CN}$) are listed in Table~\ref{tab:prodrate} on a logarithmic scale. Several measurements beyond $r_{\rm h}\sim2.5$\,au are unavailable due to low signal-to-noise ratios (SNRs).

Figure 2 illustrates the logarithmic $Q_{\rm CN}$ as a function of $r_{\rm h}$, with pre-perihelion measurements from \citet{Hutsemekers:2026} included for comparison. Over the post-perihelion range covered by our data, $r_{\rm h}\sim1.8$--3.3 au, $Q_{\rm CN}$ declines as $r_{\rm h}^{-n}$, with a weighted best-fit slope of $n = 3.5 \pm 0.2$, significantly shallower than the inbound value of $n = 6.7 \pm 0.1$.
This shallow decline is consistent with the slope of $n=3.7\pm0.6$ reported by \citet{2026arXiv260116983H}, based on an early post-perihelion measurement of $\log Q_{\rm CN}=26.2\pm0.1$ at $r_{\rm h}=1.51$\,au and two TRAPPIST measurements of $\log Q_{\rm CN}=25.74\pm0.02$ at $r_{\rm h}=1.76$\,au and $25.56\pm0.03$ at $r_{\rm h}=1.98$\,au \citep{Jehin:2025,Jehin:2025b}. Yet their trend is offset upward from ours by $\sim0.2$ dex. For the two TRAPPIST measurements, this offset is most likely dominated by their use of a constant expansion velocity of $v_{\rm exp}=1$\,km\,s$^{-1}$. The effects of different $g$-factors and scale lengths are relatively small.

A similar perihelion asymmetry was observed in the water production rate ($Q_{\rm H_2O}$) of 3I, which transitioned from a steep inbound slope ($n = 5.9 \pm 0.8$) to a significantly shallower outbound decline \citep[$n = 3.3 \pm 0.3$;][]{Tan:2026}. The comparable slopes of $Q_{\rm H_2O}$ and $Q_{\rm HCN}$ indicate a relatively stable $Q_{\rm CN}/Q_{\rm H_2O}$ mixing ratio across perihelion. Specifically, combining our measurements with the $Q_{\rm H_2O}$ data from \citet{Tan:2026} yields $Q_{\rm CN}/Q_{\rm H_2O} = 0.13$--$0.19\%$ over $r_{\rm h}\sim1.9$--2.2\,au. This is consistent with the 0.18--0.28\% range reported by \citet{Kawakita:2026} at similar heliocentric distances, although their water production rates were estimated from [O\,{\sc i}] $\lambda6300$ and are therefore subject to additional systematic uncertainty.

Since HCN is generally considered the dominant parent species of CN, measurements of HCN relative to H$_2$O provide an additional consistency check for the CN measurements. Before perihelion, JCMT observations yielded a comparable abundance of $Q_{\rm HCN}/Q_{\rm H_2O} = 0.20 \pm 0.08\%$ at 2.1\,au \citep{Coulson:2026}. However, a significantly lower value of $0.062\pm0.003\%$ was reported near perihelion using IRAM \citep{Biver:2026}. Given that the IRAM observations were obtained about one month before our first epoch, one possibility is that HCN remained the dominant parent of CN, while HCN and CN evolved differently from H$_2$O between the two epochs, resulting in a higher $Q_{\rm CN}/Q_{\rm H_2O}$ ratio at the time of our observations. Alternatively, additional parent species besides HCN may have contributed to the CN abundance after perihelion. The former possibility is hinted at by the TRAPPIST measurements spanning November 22 to December 7, which show different $r_{\rm h}$-dependences for CN and OH \citep{Jehin:2025,Jehin:2025b}. Additional observations of HCN and CN are still needed to distinguish between these scenarios. Once again, this highlights the importance of long-term, multi-epoch monitoring to trace cometary activity and compositional evolution within a self-consistent temporal framework.

% A stable mixing ratio, however, does not necessarily imply identical outgassing regions or release mechanisms. During the inbound leg, $Q_{\rm H_2O}$ included substantial contributions from icy grains in the extended coma \citep{Yang:2025,Xing:2025,Cordiner:2025,Hutsemekers:2026,Tan:2026}, whereas ALMA mapping showed that HCN production was consistent with direct nucleus sublimation \citep{Roth:2025}. 
% Similar decoupling between the outgassing regions of H$_2$O and CN has also been observed in solar system comets, despite relatively stable mixing ratios \citep{Knight:2013,Opitom:2015,Villa:2024}.
% % , indicating that stable bulk mixing ratios do not necessarily translate into uniform outgassing behavior for both solar system and interstellar comets.
% % These results point to distinct physical origins despite the broadly constant $Q_{\rm CN}/Q_{\rm H_2O}$ ratio.
% After perihelion, the gradual decline of $Q_{\rm H_2O}$ was attributed to a nearly constant effective active area regulated by decreasing solar insolation \citep{Tan:2026}, yet 
% % in the absence of CN or HCN mapping during this phase, 
% the spatial origin of our observed CN remains unconstrained.

The mixing ratios of C$_3$, C$_2$, and CH relative to CN remain generally stable, with no statistically significant trend detected over the observed heliocentric range. The weighted averages yield $\log (Q_{\rm C_3}/Q_{\rm CN})=-1.03\pm0.02$, $\log (Q_{\rm C_2}/Q_{\rm CN})=-0.08\pm0.02$, and $\log (Q_{\rm CH}/Q_{\rm CN})=0.21\pm0.02$, indicating moderate depletion in carbon-chain species yet a ``typical'' CH abundance under the compositional classification of \citet{Cochran:2012}. 

% Our derived $\log (Q_{\rm C_2}/Q_{\rm CN})$ is consistent the November 29 measuremnts -0.11+0.06-0.07reported by \citet{Kawakita:2026}, but slightly higher than their later measuments as well as the measurement of \citet{2026arXiv260116983H} -0.26+-0.14. The $\log (Q_{\rm C_3}/Q_{\rm CN})$ is also higher than \citet{2026arXiv260116983H} which gives -1.51+-0.14. Since production rate ratios are independent from the expansion velocity assumed, Such discrepency are more likely arise from different aperture used and different bandpasses. Narrowband imaging by TRAPPIST suggested more enriched values of $\log(Q_{\rm C_3}/Q_{\rm CN})\sim-0.83$ and $\log(Q_{\rm C_2}/Q_{\rm CN})\sim0.05$ at $r_{\rm h}=1.76$--1.98\,au, placing 3I even within the ``typical'' compositional class \citep{Jehin:2025,Jehin:2025b}. 

Among the published post-perihelion spectroscopic results, our $\log (Q_{\rm C_2}/Q_{\rm CN})$ is consistent with the November 29 value of $-0.11^{+0.06}_{-0.07}$ reported by \citet{Kawakita:2026}, but is higher than their later measurements and than the $-0.26\pm0.14$ reported by \citet{2026arXiv260116983H}. Our $\log (Q_{\rm C_3}/Q_{\rm CN})$ is also higher than the $-1.51\pm0.14$ reported by \citet{2026arXiv260116983H}. Since mixing ratios are independent of the adopted expansion velocity, the spread is more likely related to the use of different extraction apertures and spectral bandpasses. This difference can be even more pronounced when compared with narrow-band imaging, which typically uses a larger aperture and a different background-subtraction procedure from spectroscopy. In particular, TRAPPIST reported much less depleted values, with $\log (Q_{\rm C_3}/Q_{\rm CN})\sim-0.83$ and $\log (Q_{\rm C_2}/Q_{\rm CN})\sim0.05$ at $r_{\rm h}=1.76$--$1.98$\,au \citep{Jehin:2025,Jehin:2025b}, placing 3I even within the ``typical'' compositional class. A similarly less depleted value of $\log (Q_{\rm C_2}/Q_{\rm CN})=-0.01$ was reported by D. Schleicher at $r_{\rm h}=2.35$\,au post-perihelion \citep{Kawakita:2026}.

In their 
% restricted subsample of their 
thirty-year survey, \citet{Cochran:2012} reported a lower ratio of $\log(Q_{\rm CH}/Q_{\rm CN}) = 0.02 \pm 0.26$ for carbon-chain depleted comets, compared to $0.25 \pm 0.18$ for the ``typical'' group. However, the large dispersion and the very small depleted sample (only two with CH measurements) render the difference statistically insignificant. Inspection of individual comets in their sample further shows that comets classified as carbon-chain depleted are not consistently depleted in CH, indicating only a weak correlation between CH abundance and carbon-chain species \citep{Cochran:2012}. Our derived ratio of $\log(Q_{\rm CH}/Q_{\rm CN}) = 0.21\pm0.02$ for 3I lies near the midpoint of the ``typical'' group and is in good agreement with the long-period comet average of $0.27 \pm 0.17$ \citep{Cochran:2012}.

Proposed parents of CH observed in the cometary comae include CH$_4$ as a nucleus ice or
distributed organic compounds \citep{Cochran:1997}. Recent JWST measurements show that 3I is somewhat enriched in CH$_4$ relative to most solar system comets \citep{Roth:2026b,Belyakov:2026}. The absence of a correspondingly elevated CH abundance in our observations therefore disfavors CH$_4$ as the dominant parent of CH in 3I. 
% The mis-match between CH$_4$ and CH was also observed in comet Hale-Bopp, where models based solely on CH$_4$ could not fully reproduce the observed column densities of CH. For Hale-Bopp,  spatially resolved observations also revealed no clear correlation between CH emission and dust, leaving the origin of CH as an open question \citep{Cochran:1997}. 

\begin{deluxetable*}{lccccccccccc}
    \tablecaption{Production rates and mixing ratios.\label{tab:prodrate}}
    \tablewidth{0pt}
    \tablehead{
        \colhead{Date} & \colhead{$r_h$} & \colhead{$\log Q_{\rm CN}$} & \colhead{$\log Q_{\rm C_3}$} & \colhead{$\log Q_{\rm C_2}$} & \colhead{$\log Q_{\rm CH}$}& $\log(Q_{\rm C_{3}}/Q_{\rm CN})$ & $\log(Q_{\rm C_{2}}/Q_{\rm CN})$ & $\log(Q_{\rm CH}/Q_{\rm CN})$ & $\log Q_{\rm Ni\,I}$ & $\log Q_{\rm Fe\,I}$ & $\log(Q_{\rm Ni\,I}/Q_{\rm Fe\,I})$\\
        \colhead{(UT)} & \colhead{(au)} & \colhead{(s$^{-1}$)} & \colhead{(s$^{-1}$)} & \colhead{(s$^{-1}$)} & \colhead{(s$^{-1}$)} & \colhead{} & \colhead{} & \colhead{} & \colhead{(s$^{-1}$)} & \colhead{(s$^{-1}$)} & \colhead{}
    }
    \startdata
    Dec. 2  & 1.846 & $25.33 \pm 0.03$ & $24.27 \pm 0.02$ & $25.24 \pm 0.02$ & $25.51 \pm 0.03$ & $-1.06 \pm 0.04$ & $-0.10 \pm 0.04$ & $0.18 \pm 0.04$ & $24.50 \pm 0.02$ & $24.31 \pm 0.02$ & $0.19 \pm 0.03$ \\
    Dec. 4  & 1.898 & $25.43 \pm 0.04$ & $24.36 \pm 0.02$ & $25.35 \pm 0.02$ & $25.64 \pm 0.03$ & $-1.06 \pm 0.05$ & $-0.07 \pm 0.05$ & $0.21 \pm 0.05$ & $24.43 \pm 0.09$ & $24.30 \pm 0.09$ & $0.13 \pm 0.12$ \\
    Dec. 6  & 1.947 & $25.39 \pm 0.04$ & $24.36 \pm 0.02$ & $25.30 \pm 0.02$ & $25.61 \pm 0.03$ & $-1.02 \pm 0.05$ & $-0.09 \pm 0.05$ & $0.22 \pm 0.05$ & $24.38 \pm 0.09$ & $24.24 \pm 0.09$ & $0.14 \pm 0.12$ \\
    Dec. 7  & 1.973 & $25.39 \pm 0.04$ & $24.31 \pm 0.02$ & $25.32 \pm 0.02$ & $25.72 \pm 0.03$ & $-1.08 \pm 0.05$ & $-0.07 \pm 0.05$ & $0.33 \pm 0.05$ & $24.33 \pm 0.09$ & $24.17 \pm 0.09$ & $0.16 \pm 0.12$ \\
    Dec. 17 & 2.246 & $25.21 \pm 0.04$ & $24.03 \pm 0.02$ & $25.10 \pm 0.02$ & $25.35 \pm 0.04$ & $-1.19 \pm 0.05$ & $-0.11 \pm 0.05$ & $0.13 \pm 0.06$ & $24.18 \pm 0.09$ & $23.75 \pm 0.09$ & $0.43 \pm 0.12$ \\
    Dec. 19 & 2.306 & $25.36 \pm 0.04$ & $24.21 \pm 0.03$ & $25.24 \pm 0.03$ & $25.43 \pm 0.08$ & $-1.15 \pm 0.05$ & $-0.12 \pm 0.05$ & $0.07 \pm 0.09$ & $24.01 \pm 0.09$ & $23.61 \pm 0.09$ & $0.40 \pm 0.13$ \\
    Dec. 20 & 2.333 & $25.16 \pm 0.03$ & $24.22 \pm 0.03$ & $25.01 \pm 0.03$ & ---              & $-0.94 \pm 0.04$ & $-0.15 \pm 0.04$ & ---             & $24.13 \pm 0.02$ & $23.62 \pm 0.03$ & $0.51 \pm 0.04$ \\
    Dec. 23 & 2.423 & $25.08 \pm 0.03$ & $24.19 \pm 0.03$ & $25.12 \pm 0.02$ & ---              & $-0.89 \pm 0.04$ & $0.04 \pm 0.04$  & ---             & $24.04 \pm 0.02$ & $23.39 \pm 0.03$ & $0.64 \pm 0.04$ \\
    Dec. 26 & 2.509 & $25.03 \pm 0.03$ & ---              & ---              & ---              & ---              & ---              & ---             & $24.00 \pm 0.02$ & ---              & ---             \\
    Dec. 28 & 2.570 & $24.91 \pm 0.04$ & ---              & ---              & ---              & ---              & ---              & ---             & $23.88 \pm 0.09$ & $23.24 \pm 0.09$ & $0.64 \pm 0.13$ \\
    Jan. 11 & 3.002 & $24.68 \pm 0.04$ & ---              & ---              & ---              & ---              & ---              & ---             & $23.69 \pm 0.09$ & ---              & ---             \\
    Jan. 20 & 3.293 & $24.45 \pm 0.05$ & ---              & ---              & ---              & ---              & ---              & ---             & $23.58 \pm 0.09$ & ---              & ---             \\
    \enddata
\end{deluxetable*} 

\subsection{Iron and Nickel} \label{subsec:metal}

\begin{figure*}[ht!]
    \centering
    \includegraphics[width=0.7\textwidth]{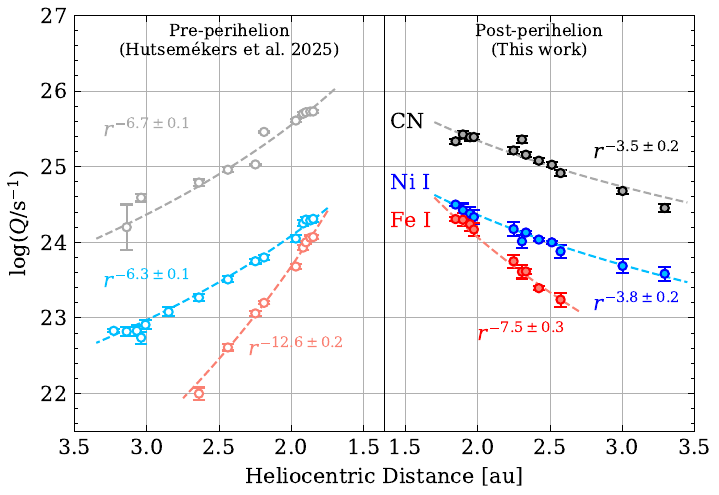}
    \caption{Production rates of CN, Ni\,{\sc i}, and Fe\,{\sc i} as a function of heliocentric distance for 3I/ATLAS. Post-perihelion measurements from this work are shown as filled circles, while pre-perihelion measurements from \citet{Hutsemekers:2026} are shown as open circles. The species are color-coded as CN---grey, Ni\,{\sc i}---blue, and Fe\,{\sc i}---red. Dashed lines indicate the corresponding power-law dependencies ($r_{\rm h}^{-n}$) derived from weighted fits.\label{fig:trend}}
\end{figure*}

To derive the production rates of Fe\,{\sc i} and Ni\,{\sc i}, we adopt the many-level fluorescence model of \citet{Bromley:2021}. This model treats the atomic system under the assumptions of a collisionless and optically thin coma, and computes line-by-line $g$-factors by solving the statistical equilibrium including spontaneous emission, stimulated emission, and absorption. In \citet{Bromley:2021}, the line-by-line $g$-factors were used to derive Fe\,{\sc i} and Ni\,{\sc i} column densities from individual emission lines, which were then averaged and converted to production rates using a Haser model. For our data, however, the low spectral resolution does not allow reliable isolation of individual lines. We therefore use the line-by-line $g$-factors to construct a synthetic low-resolution spectrum, and derive the column densities $N_{\rm Fe\,I}$ and $N_{\rm Ni\,I}$ by fitting this synthetic spectrum to the observed spectrum at each epoch. 

Specifically, the synthetic spectrum is of the form
\begin{equation}
    I^{\rm mod}_\lambda=\frac{1}{4\pi}\left[N_{\rm Fe\,I}\sum_i g_{i}\,\phi_\lambda(\lambda_i)+N_{\rm Ni\,I}\sum_jg_{j}\,\phi_\lambda(\lambda_j)\right]~~,
\end{equation}
where $I^{\rm mod}_\lambda$ is the modeled spectral intensity, $g_i$ and $\lambda_i$ (or $g_j$ and $\lambda_j$) are the $g$-factor and wavelength of the $i$-th Fe\,{\sc i} (or $j$-th Ni\,{\sc i}) transition, and $\phi_\lambda$ is the Gaussian line profile with a fixed FWHM of 15.7\,${\rm\AA}$ for the BFOSC data and 18.0\,${\rm\AA}$ for the YFOSC data. 
For transitions blueward of 3650\,${\rm \AA}$, the wavelengths are adjusted according to $\lambda' = (1 + s)\lambda -s\times3650\,{\rm \AA}$ to correct for the residual linear stretch at the blue end (see Appendix~\ref{app:data}). 
The stretch factor $s$ is treated as a free parameter in the fitting, with the final free parameter set given by $\{s,N_{\rm Fe\,I},N_{\rm Ni\,I}\}$.

The synthetic spectrum is fitted to the observed spectrum using Markov Chain Monte Carlo (MCMC) sampling with a Gaussian log-likelihood function,
\begin{equation}
    \ln \mathcal{L}=
    -\frac{1}{2}
    \sum_{\lambda}
    \left(
    \frac{I^{\rm obs}_\lambda - I^{\rm mod}_\lambda}{\sigma^{\rm obs}_\lambda}
    \right)^2~~,
\end{equation}
where $I^{\rm obs}_\lambda$ and $\sigma^{\rm obs}_\lambda$ are the observed intensity and its uncertainty at $\lambda$. To isolate the metal-emission region, the fitting is restricted to wavelengths shorter than 3840\,${\rm\AA}$. The interval 3346--3375\,${\rm\AA}$ is excluded to avoid potential contamination from the NH ($\Delta\nu=0$) band.

For each epoch, we run the sampling twice. The first run provides an initial fit to the spectra, from which we adopt the best-fit stretch factor $s$ to refine the wavelength solution at $\lambda < 3650\,{\rm\AA}$. We then repeat the flux calibration using the corrected wavelength solution (see Appendix~\ref{app:data}). The MCMC sampling is subsequently performed on the recalibrated spectra to derive the final column densities $N_{\rm Fe\,I}$ and $N_{\rm Ni\,I}$, which we convert to production rates ($Q_{\rm Fe\,I}$ and $Q_{\rm Ni\,I}$) using the Haser model adopted by \citet{Bromley:2021} and the same $v_{\rm exp}$ as for the daughter molecules. The resulting $Q_{\rm Fe\,I}$, $Q_{\rm Ni\,I}$, and the $Q_{\rm Ni\,I}/Q_{\rm Fe\,I}$ ratio are listed in Table~\ref{tab:prodrate} on a logarithmic scale. Due to the high noise level relative to the weakening emission, $Q_{\rm Fe\,I}$ could not be reliably constrained for the epochs of December 26, January 11, and 20.

The logarithmic production rates are shown in Figure~\ref{fig:trend} as a function of heliocentric distance, with pre-perihelion measurements from \citet{Hutsemekers:2026} included for comparison. 
Overall, our post-perihelion $\log Q_{\rm Fe\,I}$ and $\log Q_{\rm Ni\,I}$ are higher than the pre-perihelion values. The offset grows from only $\sim0.2$ dex (a factor of 1.5) at $r_{\rm h}\sim1.85$\,au to progressively larger values at greater $r_{\rm h}$, reflecting the different r$_{\rm h}$-dependences before and after perihelion. Specifically, in our post-perihelion data, $\log Q_{\rm Fe\,I}$ and $\log Q_{\rm Ni\,I}$ follow power-law slopes of $n=3.8\pm0.2$ and $n=7.5\pm0.3$, respectively, both significantly shallower than the corresponding inbound values of $n=6.3\pm0.1$ and $n=12.6\pm0.2$. 

% Our derived $\log Q_{\rm Ni\,I}$ is consistent with the range of 23.7--24.2 reported by \citet{Belyakov:2026} at $r_{\rm h}\sim2.2$\,au.
A comparison with the VLT/UVES measurements of \citet{Hutsemekers:2026b} shows excellent agreement in $\log Q_{\rm Fe\,I}$, whereas our $\log Q_{\rm Ni\,I}$ values are systematically higher by $\sim0.4$ dex (a factor of $\sim2$). Part of this discrepancy may arise from differences in the adopted Ni\,{\sc i} g-factors, which can vary by up to $\sim0.15$\,dex between fluorescence models because of uncertainties in the atomic data (D. Hutsem\'ekers, private comm.). The remaining discrepancy is less clear. As an independent check, applying the same fluorescence model to the KCWI spectrum of \citet{2026arXiv260116983H} yields $\log Q_{\rm Ni\,I}\sim24.80$ and $\log Q_{\rm Fe\,I}\sim24.87$ (W. B. Hoogendam, private comm.), both consistent within $\lesssim0.1$\,dex with the VLT/UVES trend extrapolated to the same $r_{\rm h}$. Given the excellent agreement in $\log Q_{\rm Fe\,I}$, the systematically higher $\log Q_{\rm Ni\,I}$ may reflect a calibration uncertainty affecting the very blue end of the spectrum, where the spectrograph throughput is low and the Ni\,{\sc i} lines are concentrated. Even a small overcorrection in the flux calibration would preferentially enhance the Ni\,{\sc i} fluxes relative to Fe\,{\sc i}. 
The comparison with \citet{Hutsemekers:2026b} further indicates that uncertainties in the Ni\,{\sc i} fluxes have little impact on the heliocentric trend. Hence, even if the Ni\,{\sc i} fluxes are systematically elevated, our key findings remain unchanged: Ni\,{\sc i} declines more slowly than Fe\,{\sc i} after perihelion, and its post-perihelion decline is also slower than during the pre-perihelion phase.

% But it is $\sim1$ dex lower than $\log Q_{\rm Ni\,I} = 25.8\pm0.2$ reported by \citet{2026arXiv260116983H} at $r_{\rm h}\sim1.51$\,au (so as $Q_{\rm Fe\,I}$). This difference is primarily due to the simplified three-level model adopted by \citet{2026arXiv260116983H}, which typically requires a blackbody temperature much lower than the solar color temperature of $\sim5800$\,K in order to match the observed level populations, but at the same time underestimates the pumping flux \citep[D. Hutsem\'ekers, private comm.;][]{Manfroid:2021}. Applying the many-level fluorescence model to their spectrum gives revised values of $\log Q_{\rm Ni\,I}\sim24.80$ and $\log Q_{\rm Fe\,I}\sim24.87$, both consistent with our derived trends to within 0.1 dex when extrapolated to the same $r_{\rm h}$ . 

The steep pre-perihelion heliocentric dependencies of $Q_{\rm Fe\,I}$ and $Q_{\rm Ni\,I}$ and the resulting evolution of the $Q_{\rm Ni\,I}/Q_{\rm Fe\,I}$ ratio were interpreted as a natural consequence of the temperature-dependent sublimation of highly volatile metal carbonyls such as Ni(CO)$_4$ and Fe(CO)$_5$ \citep{Hutsemekers:2026}. Our post-perihelion measurements, however, show significantly shallower heliocentric trends. This behavior reveals a clear perihelion asymmetry, analogous to that seen in $Q_{\rm H_2O}$ \citep{Tan:2026}, and indicates that the release of Fe\,{\sc i} and Ni\,{\sc i} is likely also regulated by the thermal evolution and progressive surface modification of the nucleus, rather than being controlled solely by the intrinsic volatility of metal carbonyls, which would otherwise produce a more symmetric heliocentric dependence across perihelion.

% A similar perihelion asymmetry was observed in the water production rate ($Q_{\rm H_2O}$), which transitioned from a steep inbound slope ($n=5.9\pm0.8$) to a significantly shallower outbound decline \citep[$n=3.3\pm0.3$;][]{Tan:2026}, and 
% The transition in $Q_{\rm H_2O}$ was interpreted as the result of a dynamically expanding active surface area before perihelion that stabilized afterward, with the post-perihelion activity decline primarily regulated by the decreasing solar insolation. 

\subsection{[O\,I] $\lambda$6300 Emission}\label{subsec:OI}

The [O\,{\sc i}] $\lambda\lambda6300,6364$ (red) doublet arises from the radiative decay of oxygen atoms in the metastable O($^1$D) state, which is primarily produced via the photodissociation of oxygen-bearing parent molecules, including H$_2$O, CO$_2$, and CO \citep{Festou:1981,McKay:2012}. The same parent molecules can also produce oxygen atoms in the O($^1$S) state, whose radiative decay gives rise to the [O\,{\sc i}] $\lambda5577$ (green) line. Since different parents contribute differently to the O($^1$S) and O($^1$D) states, the green-to-red line ratio therefore provides a useful diagnostic of CO$_2$ or CO abundance relative to H$_2$O \citep[e.g.,][]{McKay:2012,Decock:2013,Opitom:2021,Shinnaka:2026}. At our spectral resolution, however, the [O\,{\sc i}] $\lambda5577$ line cannot be cleanly separated from the adjacent C$_2$ emission band, precluding a robust green-to-red analysis.
% and thus any direct constraint on the relative abundance of CO$_2$ or CO. 
Instead, we test whether the observed [O\,{\sc i}] emission can be explained entirely by the three major oxygen-bearing parents by calculating their expected contributions from the published production rates. Similar approaches that relate the [O\,{\sc i}] line intensity to production rate of the dominant parent molecules were adopted by previous studies, e.g., \citet{Hyland:2019} and \citet{Kawakita:2026}.

In our spectra, the $\lambda6300$ component of the red doublet is well detected in the first three epochs ($r_{\rm h}=1.85$--1.97\,au), whereas $\lambda6364$ is only marginally detected (Figure~\ref{fig:spec}). We therefore retain $\lambda6300$ for further analysis and exclude $\lambda6364$ because of its low SNR. For each epoch, the wavelength-integrated intensity of $\lambda6300$, $I_{6300}$, is measured by integrating a fitted Gaussian profile. The resulting values range from 0.13 to 0.17\,erg\,s$^{-1}$\,cm$^{-2}$\,sr$^{-1}$, with relative uncertainties of 7--12\% (see Table~\ref{tab:flux}). Despite its higher SNR, $I_{6300}$ may still be affected by telluric absorption, imperfect subtraction of sky background, and possible contamination from the nearby NH$_2$ (0,8,0) band. Their importance are assessed in Appendix~\ref{app:OI}, where we find only minor telluric absorption, a sky-background bias of $\sim2$\%, and NH$_2$ contamination below 6\%, supporting the use of $I_{6300}$ in the analysis below.

To calculate the expected contributions from H$_2$O, CO$_2$, and CO, we first derive the column density $N_{\rm parent}$ of each parent molecule from the published production rates using a Haser model \citep{Haser:1957}, and then estimate the expected contribution through\footnote{Our expression differs from the original form used by \citet{Bhardwaj:2012}, because here $I$ is expressed in erg\,s$^{-1}$\,cm$^{-2}$\,sr$^{-1}$.}
\begin{equation}
    I=\frac{h\nu\,}{4\pi}\tau^{-1}\,\alpha\,\beta\,N_{\rm parent}~~, 
\end{equation}
where $h\nu$ is the photon energy at 6300\,${\rm\AA}$, $\tau$ is the photodissociative lifetime of the parent molecule, $\alpha$ is the photodissociative yield into the O($^1$D) state, and $\beta=0.75$ is the branching ratio into the $\lambda6300$ line \citep{Bhardwaj:2012}. The $\tau$ and $\alpha$ for each parent molecule are adopted from \citet{McKay:2012}, whose values are summarized in Table~\ref{tab:OI}.

For the input $Q_{\rm H_2O}$, we adopt the values of 1.30--$2.46\times10^{28}$\,s$^{-1}$ reported by \citet{Tan:2026} on nearby dates. 
For $Q_{\rm CO_2}$, we linearly extrapolate the two measurements of \citet{Belyakov:2026} in the $\log Q_{\rm CO_2}$--$\log r_{\rm h}$ space to the $r_{\rm h}$ of our observations, yielding 1.26--$1.49\times10^{28}$\,s$^{-1}$. We note that the original \citet{Belyakov:2026} measurements are more than three times higher than the \citet{Cordiner:2026} value at similar $r_{\rm h}$ (still twice higher after correcting for the different $v_{\rm exp}$), and are therefore a conservative choice.
For $Q_{\rm CO}$, we adopt the perihelion value of $6.8\times10^{27}$\,s$^{-1}$ reported by \citet{Biver:2026} as an upper limit.
The expansion velocity $v_{\rm exp}$ for each parent molecule is taken from the same reference as the production rates, with the adopted values summarized in Table~\ref{tab:OI}. 

We calculate for the first three epochs and find that H$_2$O accounts for 13--22\% of the observed $I_{6300}$, whereas CO$_2$ and CO contribute 29--31\% and 4--5\%, respectively. A residual of 44--53\% therefore remains, well beyond the flux uncertainties and the estimated telluric absorption, sky-subtraction bias, and NH$_2$ contamination discussed above, indicating that additional sources of systematic uncertainty need to be considered. 

A major uncertainty could arise from the photodissociative yields, $\alpha$, of the parent molecules. By definition, $\alpha$ is the solar-flux-weighted average of the wavelength-dependent yield $\alpha(\lambda)$ over the UV wavelengths capable of producing O($^1$D), and should ideally be based on wavelength-resolved laboratory measurements \citep{Huebner:1992}. In practice, however, such measurements remain sparse. Commonly used compilations of $\alpha$ rely partly on assumed or extrapolated values of $\alpha(\lambda)$ at unmeasured wavelengths, potentially introducing substantial systematic uncertainties \citep{Bhardwaj:2012}. For example, based on the observed [O\,{\sc i}] green-to-red ratios of C/2016 R2, \citet{Raghuram:2020} suggested that $\alpha_{\rm CO_2}$ could be underestimated by a factor of three. Although increasing $\alpha_{\rm CO_2}$ threefold would substantially reduce the residual, determining whether this correction applies to 3I would still require a reliable green-to-red ratio, which cannot be measured from our low-resolution spectra. Additional uncertainty could also arise from the omission of non-photodissociative processes, such as electron-impact dissociative excitation, although their contributions are small compared with photodissociation \citep{Bhardwaj:2012,Raghuram:2020} and are unlikely to account for the observed residual.

% For H$_2$O, the commonly used values \citep[including that from][]{McKay:2012} are based on the compilation by \citet{Huebner:1992} which is anchored by only two measurements at 1216 and 1236\,${\rm \AA}$, with assumed yields adopted over the remaining wavelength ranges. For CO$_2$, no direct measurement of $\alpha(\lambda)$ exists over the relevant wavelength range. The commonly adopted value is instead derived from ...
% Indeed, . 

Another possible explanation for the substantial residual is a contribution from additional oxygen-bearing parent species. For example, the discovery of O$_2$ in the coma of 67P/Churyumov--Gerasimenko, together with the reanalysis of Giotto data for 1P/Halley, suggests that O$_2$ may be common in solar system comets \citep{Bieler:2015,Rubin:2015}. Models suggest that cometary O$_2$ can be primordial or formed before comet accretion and subsequently trapped in icy grains or clathrates \citep{Mousis:2016}. It could therefore also be present in an ISO such as 3I and contribute to the observed [O\,{\sc i}] emission. According to \citet{Huebner:1992}, O($^1$D) is a major photodissociation product of O$_2$, with an effective yield of $\alpha_{\rm O_2}\sim0.84$. Thus, even an O$_2$ abundance of only a few percent could provide a non-negligible contribution to the $\lambda6300$ emission. Unfortunately, without a direct measurement of O$_2$ in 3I, its actual contribution cannot be quantified in this work.

\section{Discussion} \label{sec:dis}

\begin{figure*}[ht!]
    \centering
    % 第一张图
    \begin{subfigure}[b]{0.325\textwidth}
        \centering
        \includegraphics[width=\textwidth]{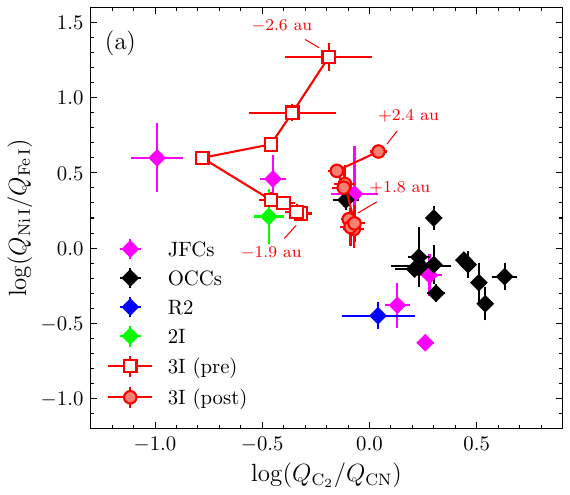}
    \end{subfigure}
    \begin{subfigure}[b]{0.325\textwidth}
        \centering
        \includegraphics[width=\textwidth]{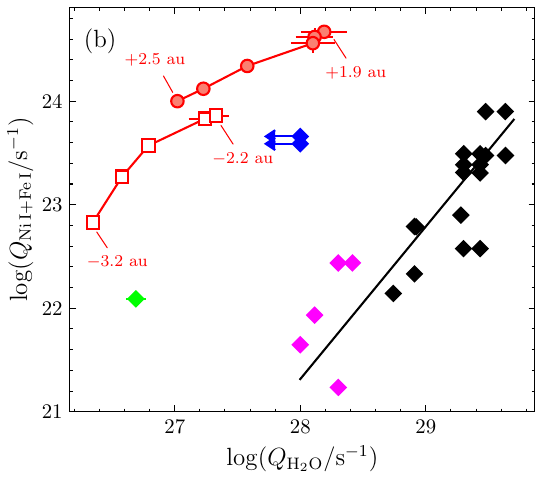}
    \end{subfigure}
    \begin{subfigure}[b]{0.325\textwidth}
        \centering
        \includegraphics[width=\textwidth]{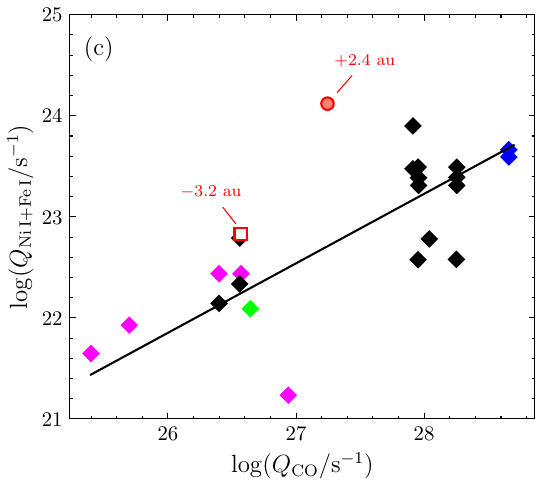}
    \end{subfigure}
    \caption{
    Correlations. (a): $\log(Q_{\rm Ni\,I}/Q_{\rm Fe\,I})$ versus $\log(Q_{\rm C_2}/Q_{\rm CN})$; (b): $\log(Q_{\rm Ni\,I+Fe\,I})$ versus $\log(Q_{\text{H}_2\text{O}})$; (c): $\log(Q_{\rm Ni\,I+Fe\,I})$ versus $\log(Q_{\text{CO}})$. Selected points are labeled by heliocentric distance to illustrate the temporal evolution of 3I, with negative and positive values denoting pre- and post-perihelion, respectively. The black lines in panel (b) and (c) denote the best-fit correlations with $Q_{\text{H}_2\text{O}}$ (including all solar system comets except C/2016 R2) and with $Q_{\text{CO}}$ (including all comets). Post-perihelion measurements of 3I (red circles) are from this work, except for $Q_{\rm H_2O}$ from \citet{Tan:2026}, \citet{Cordiner:2026}, and \citet{Belyakov:2026}, and $Q_{\rm CO}$ from \citet{Cordiner:2026}. Pre-perihelion measurements of 3I (open squares) are taken from \citet{Hutsemekers:2026} and \citet{Cordiner:2025}. Comparison data are from \citet{Hutsemekers:2021}, \citet{Manfroid:2021}, and \citet{Opitom:2021}, with colors and marker styles following \citet{Manfroid:2021} for consistency.
    \label{fig:corr}
    }
\end{figure*}

Early pre-perihelion spectroscopy classified 3I as strongly carbon-chain depleted, with $\log(Q_{\rm C_2}/Q_{\rm CN})<-0.86$ \citep{SalazarManzano:2025,2026ApJ...998L..30L}. Later pre-perihelion measurements by \citet{Hutsemekers:2026} indicated a less extreme depletion, with $\log(Q_{\rm C_2}/Q_{\rm CN})$ values clustering around -0.4. By contrast, post-perihelion measurements span $\log(Q_{\rm C_2}/Q_{\rm CN})=-0.26$--0.05, consistent with a substantially reduced degree of C$_2$ depletion after perihelion. This change may reflect delayed release of its parent species from relatively unprocessed subsurface material, or compositional heterogeneity in the nucleus combined with seasonal illumination effects around perihelion. A similar interpretation was proposed by \citet{Jehin:2025} to explain the TRAPPIST measurements and by \citet{Belyakov:2026} to account for the enhanced CH$_4$ abundance observed after perihelion.

A statistically significant correlation between $\log(Q_{\rm C_2}/Q_{\rm CN})$ and $\log(Q_{\rm Ni\,I}/Q_{\rm Fe\,I})$ was reported by \citet{Hutsemekers:2021}  across both solar system comets and 2I/Borisov (see Figure~\ref{fig:corr}a), from which it was inferred that, similar to $\log(Q_{\rm C_2}/Q_{\rm CN})$, $\log(Q_{\rm Ni\,I}/Q_{\rm Fe\,I})$ is also somehow linked to primordial formation conditions \citep{Hutsemekers:2021}. However, complexity arises when the inbound measurements of 3I from \citet{Hutsemekers:2026} and outbound measurements from this work are added to the figure. While most comets in the original sample were observed only once or over a narrow range of $r_{\rm h}$ and thus appear as static points, multi-epoch observations of 3I span a broader range of $r_{\rm h}$ and reveal a displacement across the correlation trend. This behavior raises the question of how to define a representative value of $\log(Q_{\rm Ni\,I}/Q_{\rm Fe\,I})$ for individual objects. \citet{Hutsemekers:2026} noted that $\log(Q_{\rm Ni\,I}/Q_{\rm Fe\,I})$ approaches a stable value of $\sim0.3$ at $r_{\rm h}\lesssim2$\,au, which fits the correlation well. Using the same criterion, our first four measurements yield $\log(Q_{\rm Ni\,I}/Q_{\rm Fe\,I}) \sim 0.2$. Combined with the less depleted $\log(Q_{\rm C_2}/Q_{\rm CN})$, this representative value also follows the same correlation. However, \citet{2026arXiv260116983H} reported a much lower $\log(Q_{\rm Ni\,I}/Q_{\rm Fe\,I})=-0.16\pm0.03$ at $r_{\rm h}=1.509$\,au, suggesting that the relative enhancement of $Q_{\rm Fe\,I}$ persists well within $r_{\rm h}\lesssim2$\,au. Additional measurements at $r_{\rm h}\lesssim2$\,au will be helpful to confirm whether a truly stable $\log(Q_{\rm Ni\,I}/Q_{\rm Fe\,I})$ is reached.

% The $\log(Q_{\rm C_2}/Q_{\rm CN})$ ratios measured in solar system comets are often considered to reflect primordial formation conditions rather than evolutionary effects, such as repeated thermal processing \citep{2015SSRv..197....9C,Cochran:2020}. Evidences include the ``typical'' carbon-chain composition maintained by 2P/Encke despite its frequent perihelion passages \citep{Cochran:2012}, the identical ``depleted'' composition observed across the fragmented nucleus of 73P/Schwassmann-Wachmann 3 \citep{Schleicher:2011}, and the ``depleted'' compositions found among dynamically new comets \citep[e.g.,][]{Cambianica:2025}.

% However, how these metal carbonyls form and are retained in the nucleus remains unknown. A possible explaination. implication for formation. 

% Compared to the relatively stable mixing ratio of $Q_{\rm CO_2}/Q_{\rm H_2O}$ \citep{Cordiner:2025,Belyakov:2026}, our results suggest that CO production is decoupled from these two species, at least during the outbound. 

On the other hand, any physical interpretation of the production rates and relative abundances of Fe\,{\sc i} and Ni\,{\sc i} ultimately depends on the nature of their parent carriers. The detection of Fe\,{\sc i} and Ni\,{\sc i} far from the sun indicates the existence of volatile metal-bearing compounds in the nucleus, as refractory dust grains cannot sublimate at the low temperatures characteristic of these distances \citep{Manfroid:2021}. This is further supported by the observation that, in solar system comets, $\log(Q_{\rm Ni\,I}/Q_{\rm Fe\,I})$ measured far from the sun is significantly higher than the bulk values inferred from Sun-grazing comets, which exhibit solar abundance \citep{Manfroid:2021,Hutsemekers:2021}. 

One leading hypothesis for the volatile parents is metal carbonyls, initially proposed by \citet{Manfroid:2021} based on a stronger correlation of the metal production rates with CO than with H$_2$O (see Figure~\ref{fig:corr}b and \ref{fig:corr}c). This distinction is most clearly illustrated by the chemically peculiar comet C/2016 R2 (PanSTARRS), which, despite being remarkably water-poor, shows simultaneously high production rates of CO and of gaseous Fe\,{\sc i} and Ni\,{\sc i} \citep{Biver:2018,McKay:2019,Manfroid:2021}. A similar pattern is also seen in the water-poor interstellar comet 2I/Borisov \citep{Opitom:2021}.

As shown in Figure~\ref{fig:corr}b and \ref{fig:corr}c, 3I appears to follow the same overall picture, in that it is systematically metal-rich for its H$_2$O output while lying much closer to the metal--CO relation. The recurrence of this pattern in both chemically peculiar comets and comets of different origins points to at least a partial decoupling of the metals and CO from H$_2$O, consistent with the metals being linked to a CO-bearing volatile reservoir, potentially in the form of metal carbonyls. At the same time, the varying degrees of metal enrichment among solar system and interstellar comets suggest that the metal abundance relative to H$_2$O, and perhaps also CO abundance, may retain information on differences in the formation environment.

Finally, the temporal evolution of 3I in Figure~\ref{fig:corr}b is also noteworthy. It does not follow the same slope as the metal--H$_2$O relation defined by solar system comets, suggesting that the comet-to-comet trend is not simply represent the time evolution of individual comets. Unfortunately, in the absence of comparable multi-epoch measurements for other comets, it remains unclear whether this difference reflects a genuine distinction between interstellar and solar system comets or a more general property of cometary metal release.

\section{Summary} \label{sec:sum}
In this work, we present multi-epoch post-perihelion optical spectroscopy of 3I/ATLAS and analyze the evolution of its coma composition. Our main findings are summarized as follows:
\begin{enumerate}
    \item The post-perihelion production rate of CN declines with a  power-law index of $n=3.5\pm 0.2$, significantly shallower than the steep pre-perihelion slope of $n=6.7\pm 0.1$. This asymmetry closely resembles that of H$_2$O, yielding a relatively stable mixing ratio of $Q_{\rm CN}/Q_{\rm H_2O}\sim0.13$--0.19\% over $r_{\rm h}\sim1.9$--2.2\,au.
    \item The mixing ratios of C$_3$, C$_2$, and CH relative to CN remain generally stable over the observed heliocentric range. C$_2$ shows a less depleted value of $\log (Q_{\rm C_2}/Q_{\rm CN})=-0.08\pm0.02$ compared to pre-perihelion, suggesting either delayed release from relatively unprocessed subsurface layers or compositional heterogeneity influenced by seasonal illumination.
    \item  The outbound production rates of Fe\,{\sc i} and Ni\,{\sc i} are broadly comparable to the inbound values near perihelion, but become increasingly higher at larger $r_{\rm h}$, reflecting a different $r_{\rm h}$-dependence. Their asymmetric heliocentric behavior following H$_2$O suggests that gaseous metal release is not solely driven by the intrinsic volatility of the parent species, but also regulated by the progressive thermal evolution and surface modification of the nucleus.
    \item The [O\,{\sc i}] $\lambda6300$ emission shows a statistically significant residual after subtracting the expected contributions from H$_2$O, CO$_2$, and CO, which may reflect systematic uncertainties in the adopted photodissociative yields or contributions from additional O-bearing parent molecules.
    \item 3I shows elevated metal production rates relative to its H$_2$O output, yet lies closer to the empirical correlations between $Q_{\rm Ni\,I+Fe\,I}$ and $Q_{\rm CO}$ defined by solar-system comets \citep{Manfroid:2021}. This suggests that gaseous metal release is linked more closely to a CO-bearing volatile reservoir, potentially in the form of metal carbonyls.
\end{enumerate}

%% Please use the acknowledgment and contribution environments. This will 
%% be anonomyized when the "anonymous" style option is used. 
\begin{acknowledgments}
This work is supported by the National Natural Science Foundation of China (NSFC) through the projects 12588202. X.L.Z and X.Y.F acknowledge Yunnan Fundamental Research Project (Grant No.~202501AS070006). S.W. acknowledge support from the Youth Innovation Promotion Association of the CAS with No.~2023065. We acknowledge the support of the staff of the Xinglong 2.16m telescope. This work was partially supported by National Astronomical Observatories, Chinese Academy of Sciences. We acknowledge the support of the staff of the Lijiang 2.4m telescope. Funding for the telescope has been provided by Chinese Academy of Sciences and the People's Government of Yunnan Province. We thank the anonymous reviewer for critical reading and constructive suggestions that help us improve this manuscript.
\end{acknowledgments}

\begin{contribution}
%%This section gives authors the space to recognize author contributions. The text inside this environment is NOT counted towards the total word quanta. At a minimum, manuscripts are expected to include this text:

R.N.Z. conducted the BFOSC observations, reduced the BFOSC and YFOSC data, performed the spectral modeling, and drafted substantial portions of the manuscript. X.L.Z. contributed to the YFOSC observations, supervised the graduate student, and reviewed the manuscript. B.Y. contributed to the development of the forward-modeling method used in the Fe\,{\sc i} and Ni\,{\sc i} analysis and a critical review of the manuscript. X.Y.F. contributed to the YFOSC observations. S.W. and J.F.L. secured funding and reviewed the manuscript. Y.H. reviewed the manuscript.

% Contributed to the Gemini observations planning and configuration, wrote part of the manuscript, and contributed to the manuscript review.

%% But authors are expected to provide more specific details, e.g. 
%%
%%SC was responsible for writing and submitting the manuscript.
%%WWM came up with the initial research concept and edited the manuscript.
%%OTS obtained the funding and edited the manuscript.
%%EBF provided the formal analysis and validation. He also edited the manuscript.
%%GEH Supervised the undergraduates, wrote the software and administers the project github and Zenodo repositories.
%%
%% Authors can use the Contributor Role Taxonomy (CRediT) at
%% for ideas on how write a good statement tailored to their needs.

\end{contribution}

\facilities{Beijing: 2.16m, YAO: 2.4m (YFOSC)}

\software{astropy \citep{2013A&A...558A..33A,2018AJ....156..123A,2022ApJ...935..167A}, 
          astroquery \citep{Ginsburg:2019},
          astro-wcpy \citep{2023ascl.soft11001Z}, 
          ccdproc \citep{Craig:2015},
          emcee \citep{Foreman-Mackey:2013},
          sbpy \citep{2019JOSS....4.1426M}
          spectres \citep{Carnall:2017}, 
          specutils \citep{Astropy-SpecutilsDevelopmentTeam:2019},
          }


\appendix

\section{Data Reduction} \label{app:data}

For both BFOSC and YFOSC observations, the data were reduced with the {\sc astro-plpy}\footref{fn:plpy} package \citep{Zhao:2026} following standard procedures for long-slit spectroscopy. All science frames were bias-subtracted, flat-field corrected, and cleaned of cosmic rays using either median combination or the L.A.Cosmic algorithm \citep{2001PASP..113.1420V}. Geometric distortion was characterized by fitting the curved arc lines with B-spline functions, after which the science frames were resampled to make each wavelength straight over the spatial direction. At each wavelength, the sky background was modeled with a linear fit to two apertures defined symmetrically about the comet trace. The distance was set to $\sim140\arcsec$ on each side, corresponding to the largest usable separation for minimizing contamination from extended gas emission, while still preserving a two-sided sky estimate. In practice, these apertures were entirely source-free at most wavelengths but could still include gas emission in the first few epochs. On December 6, for example, the CN emission could be traced to $\sim200\arcsec$ on one side (since the comet was not exactly centered along the slit) before reaching a source-free region, well beyond the adopted sky apertures. Nevertheless, the two-sided aperture definition was still retained over a one-sided, entirely source-free aperture for its robustness against local background variations and consistency across wavelengths. This choice led to a slight over-subtraction of the gas emission, corresponding to flux losses of $<3$\% for CN and $<8$\% for C$_2$ in the first four epochs. After sky subtraction, wavelength and flux calibrations were performed to produce  fully reduced two-dimensional spectra. One-dimensional spectra were then extracted using a fixed on-sky aperture of $22\arcsec$, corresponding to 80 pixels for BFOSC and 78 pixels for YFOSC.

The flux calibration for the YFOSC observations required additional considerations. Owing to scheduling constraints, 3I and the flux standard Feige 56 were observed at noticeably different airmasses on several December nights (see Table~\ref{tab:full}). By contrast, the solar analog BD+00 2717 on those epochs was observed at airmasses better matched to that of 3I. To minimize differential atmospheric effects and to ensure internal consistency among epochs, we therefore used BD+00 2717 to calibrate all December spectra of 3I.
Specifically, we first selected the December 7 spectrum as the reference spectrum of BD+00 2717, because among all December observations its synthetic magnitudes (derived by convolution with photometric filter curves) show the best agreement with the catalog magnitudes, indicating the most reliable absolute flux level and continuum color slope.
For each of the other December nights, the instrumental response was derived nightly by comparing the instrumental spectrum of BD+00 2717 on that night with the reference spectrum. The resulting response function was then applied to the comet spectrum from the same night.
For the two January observations, a different solar analog was observed owing to the rapid apparent motion of 3I on the sky, and the calibration in those cases followed standard procedures.

During the subsequent spectral analysis, we found that the wavelength accuracy in the blue region is limited by the available reference lines. For BFOSC, the Fe-Ar arc lamp provides lines down to $\sim3500\,{\rm\AA}$, whereas for YFOSC the planetary nebula NGC~2392 offers lines only to $\sim3700\,{\rm\AA}$, leaving significant residual deviations at blue end. Typical deviations reach $\sim7\,{\rm\AA}$ at $3300\,{\rm\AA}$ for BFOSC and $\sim25\,{\rm\AA}$ for YFOSC. This directly affects the analysis of Fe\,{\sc i} and Ni\,{\sc i} emission features. To account for this, a simple linear stretch correction was incorporated into the spectral modeling (see \S\ref{subsec:metal}). After applying the derived stretch correction to the wavelength solution, we repeated the flux calibration to obtain a more accurate absolute flux scale.

\startlongtable
\begin{deluxetable*}{lrcccccccrr}
    \tablenum{A1}
    \tablecaption{Full observation log of 3I/ATLAS.\label{tab:full}}
    \tablewidth{0pt}
    \tablehead{
        \colhead{Date of 2025} & \colhead{Object} & \colhead{$\Delta$\tablenotemark{a}} & \colhead{$r_h$\tablenotemark{b}} & \colhead{$\dot{r}_h$\tablenotemark{c}} & \colhead{$\alpha$\tablenotemark{d}} & \colhead{Tel./Inst.} & \colhead{$\lambda$\tablenotemark{e}} & \colhead{Slit Width} & \colhead{Exposure} & \colhead{Airmass}\\
        \colhead{(UT)} & \colhead{} & \colhead{(au)} & \colhead{(au)} & \colhead{(km\,s$^{-1}$)} & \colhead{($^{\circ}$)} & \colhead{} & \colhead{(nm)} & \colhead{($\arcsec$)} & \colhead{(s)} & \colhead{}
    }
    \startdata
    Dec. 2  & 3I/ATLAS   & 1.892 & 1.846 & 42.44 & 30.5 & 2.16-m/BFOSC & 330--660 & 2.3      & 4$\times$300 & 1.72--1.61\\
            % & BD+07 2415 &       &       &       &      & $\cdots$     & $\cdots$ & $\cdots$ & 3$\times$180\\
            & HR 4468    &       &       &       &      & $\cdots$     & $\cdots$ & $\cdots$ & 2            & 1.71\\
            & HD 107072  &       &       &       &      & $\cdots$     & $\cdots$ & $\cdots$ & 180          & 1.74\\
    \hline 
    Dec. 4  & Hilt 102   &       &       &       &      & 2.40-m/YFOSC & 320--930 & 2.5      & 180          & 1.25\\
            & BD+00 2717 &       &       &       &      & $\cdots$     & $\cdots$ & $\cdots$ & 60           & 1.37\\
            & 3I/ATLAS   & 1.872 & 1.898 & 43.71 & 30.3 & $\cdots$     & $\cdots$ & $\cdots$ & 300          & 1.21\\
    \hline
    Dec. 6  & 3I/ATLAS   & 1.855 & 1.947 & 44.78 & 29.9 & 2.40-m/YFOSC & 320--930 & 2.5      & 4$\times$300 & 1.78--1.60\\
            & BD+00 2717 &       &       &       &      & $\cdots$     & $\cdots$ & $\cdots$ & 60           & 1.36\\
            & Feige 56   &       &       &       &      & $\cdots$     & $\cdots$ & $\cdots$ & 90           & 1.53\\
    \hline
    Dec. 7  & 3I/ATLAS   & 1.846 & 1.973 & 45.32 & 29.6 & 2.40-m/YFOSC & 320--930 & 2.5      & 3$\times$300 & 1.47--1.40\\
            & BD+00 2717 &       &       &       &      & $\cdots$     & $\cdots$ & $\cdots$ & 60           & 1.27\\
            & Feige 56   &       &       &       &      & $\cdots$     & $\cdots$ & $\cdots$ & 90           & 1.38\\
    \hline
    Dec. 17 & 3I/ATLAS   & 1.798 & 2.246 & 49.48 & 25.2 & 2.40-m/YFOSC & 320--930 & 2.5      & 2$\times$300 & 1.45--1.41\\
            & BD+00 2717 &       &       &       &      & $\cdots$     & $\cdots$ & $\cdots$ & 60           & 1.50\\
            & Feige 56   &       &       &       &      & $\cdots$     & $\cdots$ & $\cdots$ & 90           & 1.79\\
    \hline
    Dec. 19 & 3I/ATLAS   & 1.798 & 2.306 & 50.15 & 23.9 & 2.40-m/YFOSC & 320--930 & 2.5      & 2$\times$300 & 1.07\\
            & NGC 2392   &       &       &       &      & $\cdots$     & $\cdots$ & $\cdots$ & 60           & 1.30\\
            & BD+00 2717 &       &       &       &      & $\cdots$     & $\cdots$ & $\cdots$ & 60           & 1.13\\
            & Feige 56   &       &       &       &      & $\cdots$     & $\cdots$ & $\cdots$ & 90           & 1.12\\
    \hline
    Dec. 20 & 3I/ATLAS      & 1.799 & 2.333 & 50.43 & 23.3 & 2.16-m/BFOSC & 330--660 & 2.3      & 3$\times$300 & 1.22--1.21\\
            & TYC 274-696-1 &       &       &       &      & $\cdots$     & $\cdots$ & $\cdots$ & 180          & 1.35\\
            & Feige 56      &       &       &       &      & $\cdots$     & $\cdots$ & $\cdots$ & 600          & 1.26\\
    \hline
    Dec. 23 & 3I/ATLAS      & 1.807 & 2.423 & 51.28 & 21.1 & 2.16-m/BFOSC & 330--660 & 2.3      & 8$\times$300 & 1.18--1.22\\
            & TYC 274-696-1 &       &       &       &      & $\cdots$     & $\cdots$ & $\cdots$ & 3$\times$180 & 1.26\\
            & Feige 56      &       &       &       &      & $\cdots$     & $\cdots$ & $\cdots$ & 600          & 1.14\\
    \hline
    Dec. 26 & 3I/ATLAS      & 1.823 & 2.509 & 52.00 & 19.0 & 2.16-m/bFOSC & 330--660 & 2.3      & 8$\times$300 & 1.19--1.26\\
            & TYC 274-696-1 &       &       &       &      & $\cdots$     & $\cdots$ & $\cdots$ & 3$\times$180 & 1.25--1.26\\
            & Feige 56      &       &       &       &      & $\cdots$     & $\cdots$ & $\cdots$ & 600          & 1.14\\
    \hline
    % Dec. 21 & 3I/ATLAS   & 1.801 & 2.366 & 50.76 & 22.5 & 2.40-m/YFOSC & G3       & 2.5      & 2$\times$300 & 1.09\\
    %         & NGC 2392   &       &       &       &      & $\cdots$     & $\cdots$ & $\cdots$ & 60           & 1.87\\
    %         & BD+00 2717 &       &       &       &      & $\cdots$     & $\cdots$ & $\cdots$ & 60           & 1.15\\
    %         & Feige 56   &       &       &       &      & $\cdots$     & $\cdots$ & $\cdots$ & 90           & 1.04\\
    % \hline
    % Dec. 24 & 3I/ATLAS   & 1.811 & 2.451 & 51.52 & 20.4 & 2.40-m/YFOSC & G3       & 2.5      & 300          & 1.15\\
    %         & NGC 2392   &       &       &       &      & $\cdots$     & $\cdots$ & $\cdots$ & 60           & 1.07\\
    %         & BD+00 2717 &       &       &       &      & $\cdots$     & $\cdots$ & $\cdots$ & 60           & 1.28\\
    %         & Feige 56   &       &       &       &      & $\cdots$     & $\cdots$ & $\cdots$ & 60           & 1.41\\
    % \hline
    % Dec. 26 & 3I/ATLAS   & 1.823 & 2.509 & 52.00 & 19.0 & 2.40-m/YFOSC & G3       & 2.5      & 2$\times$300 & 1.22--1.20\\
    %         & NGC 2392   &       &       &       &      & $\cdots$     & $\cdots$ & $\cdots$ & 60           & 1.02\\
    %         & BD+00 2717 &       &       &       &      & $\cdots$     & $\cdots$ & $\cdots$ & 60           & 1.45\\
    %         & Feige 56   &       &       &       &      & $\cdots$     & $\cdots$ & $\cdots$ & 90           & 1.71\\
    % \hline
    Dec. 28 & 3I/ATLAS   & 1.840 & 2.570 & 52.45 & 17.4 & 2.40-m/YFOSC & 320--930 & 2.5      & 300          & 1.13\\
            & NGC 2392   &       &       &       &      & $\cdots$     & $\cdots$ & $\cdots$ & 60           & 1.04\\
            & BD+00 2717 &       &       &       &      & $\cdots$     & $\cdots$ & $\cdots$ & 60           & 1.34\\
            & Feige 56   &       &       &       &      & $\cdots$     & $\cdots$ & $\cdots$ & 90           & 1.51\\
    \hline
    % Dec. 31 & 3I/ATLAS   & 1.871 & 2.661 & 53.05 & 15.1 & 2.40-m/YFOSC & G3       & 2.5      & 4$\times$300 & 1.16--1.11\\
    %         & NGC 2392   &       &       &       &      & $\cdots$     & $\cdots$ & $\cdots$ & 60           & 1.03\\
    %         & BD+00 2717 &       &       &       &      & $\cdots$     & $\cdots$ & $\cdots$ & 60           & 1.38\\
    %         & Feige 56   &       &       &       &      & $\cdots$     & $\cdots$ & $\cdots$ & 90           & 1.58\\
    % \hline
    % Jan. 8  & GD 50      &       &       &       &      & $\cdots$     & $\cdots$ & $\cdots$ & 180          & 1.45\\
    %         & 3I/ATLAS   & 1.999 & 2.907 & 54.36 &  9.0 & 2.40-m/YFOSC & G3       & 2.5      & 2$\times$300 & 1.27--1.25\\
    %         & NGC 2392   &       &       &       &      & $\cdots$     & $\cdots$ & $\cdots$ & 60           & 1.02\\
    %         & BD+00 2717 &       &       &       &      & $\cdots$     & $\cdots$ & $\cdots$ & 60           & 2.21\\
    % \hline
    Jan. 11\tablenotemark{f} & 3I/ATLAS   & 2.064 & 3.002 & 54.76 &  6.8 & 2.40-m/YFOSC & 320--930 & 2.5      & 2$\times$300 & 1.13--1.11\\
            & NGC 2392   &       &       &       &      & $\cdots$     & $\cdots$ & $\cdots$ & 60           & 1.01\\
            & HR 3454    &       &       &       &      & $\cdots$     & $\cdots$ & $\cdots$ & 2            & 1.15\\
            & HD 89010   &       &       &       &      & $\cdots$     & $\cdots$ & $\cdots$ & 6            & 1.26\\
    \hline
    % Jan. 13 & 3I/ATLAS   & 2.111 & 3.064 & 55.00 &  5.4 & 2.40-m/YFOSC & G3       & 2.5      & 2$\times$300 & 1.18--1.17\\
    %         & NGC 2392   &       &       &       &      & $\cdots$     & $\cdots$ & $\cdots$ & 60           & 1.02\\
    %         % & HD 89010   &       &       &       &      & $\cdots$     & $\cdots$ & $\cdots$ & 6            & 1.40\\
    %         & G163-50    &       &       &       &      & $\cdots$     & $\cdots$ & $\cdots$ & 120          & 1.26\\
    % \hline
    Jan. 20\tablenotemark{f} & 3I/ATLAS   & 2.311 & 3.293 & 55.75 &  1.2 & 2.40-m/YFOSC & 320--930 & 2.5      & 600          & 1.27\\
            & HD 89010   &       &       &       &      & $\cdots$     & $\cdots$ & $\cdots$ & 5            & 1.02\\
            & GD108      &       &       &       &      & $\cdots$     & $\cdots$ & $\cdots$ & 120          & 1.28\\
            % & G163-50    &       &       &       &      & $\cdots$     & $\cdots$ & $\cdots$ & 120          & 1.20\\
    % \hline
    % Jan. 25 & 3I/ATLAS   & 2.467 & 3.448 & 56.16 &  1.7 & 2.40-m/YFOSC & G3       & 2.5      & 600          & 1.10\\
    %         & NGC 2392   &       &       &       &      & $\cdots$     & $\cdots$ & $\cdots$ & 60           & 1.03\\
    %         & HD 89010   &       &       &       &      & $\cdots$     & $\cdots$ & $\cdots$ & 6            & 1.50\\
    %         & GD108      &       &       &       &      & $\cdots$     & $\cdots$ & $\cdots$ & 120          & 1.87\\
    \enddata
    \tablenotetext{a}{Geocentric distance.}
    \tablenotetext{b}{Heliocentric distance.}
    \tablenotetext{c}{Heliocentric radial velocity.}
    \tablenotetext{d}{Phase angle.}
    \tablenotetext{e}{Wavelength range.}
    \tablenotetext{f}{Date of 2026.}
    % \tablenotetext{d}{Grism or broadband filter and integration time in seconds times the number of images.}
\end{deluxetable*}

\clearpage

\section{Band-integrated Intensities and Parameters for Haser Modeling} \label{app:prodrate}

Table~\ref{tab:flux} lists the measured band-integrated intensities for the molecular bands and [O\,{\sc i}] $\lambda6300$ line, Table~\ref{tab:para} summarizes the fluorescence efficiencies, expansion velocity, and scale lengths adopted for Haser modeling, and Table~\ref{tab:OI} gives the photodissociative yields, lifetimes, and expansion velocities used to model the [O\,{\sc i}] $\lambda6300$ emission from H$_2$O, CO$_2$, and CO.

\begin{deluxetable*}{lcccccc}
    \tablenum{B1}
    \tablecaption{Band-integrated intensities.\label{tab:flux}}
    \tablewidth{0pt}
    \tablehead{
        \colhead{Date} & \colhead{$r_h$} & \colhead{$I_{\rm CN}\times10^4$} & \colhead{$I_{\rm C_{3}}\times10^4$} & \colhead{$I_{\rm C_{2}}\times10^4$} & \colhead{$I_{\rm CH}\times10^4$} & \colhead{$I_{\rm 6300}\times10^4$}\\
        \colhead{(UT)} & \colhead{(au)} & \colhead{(erg\,s$^{-1}$\,cm$^{-2}$\,sr$^{-1}$)} & \colhead{(erg\,s$^{-1}$\,cm$^{-2}$\,sr$^{-1}$)} & \colhead{(erg\,s$^{-1}$\,cm$^{-2}$\,sr$^{-1}$)} & \colhead{(erg\,s$^{-1}$\,cm$^{-2}$\,sr$^{-1}$)} & \colhead{(erg\,s$^{-1}$\,cm$^{-2}$\,sr$^{-1}$)}
    }
    \startdata
    Dec. 2  & 1.846 & $4.2\pm0.3$   & $1.26\pm0.07$ & $2.01\pm0.10$ & $0.26\pm0.02$ & $0.13\pm0.01$\\
    Dec. 4  & 1.898 & $4.1\pm0.4$   & $1.07\pm0.06$ & $1.99\pm0.10$ & $0.25\pm0.02$ & $0.17\pm0.02$\\
    Dec. 6  & 1.947 & $3.6\pm0.4$   & $1.00\pm0.05$ & $1.62\pm0.08$ & $0.22\pm0.01$ & $0.15\pm0.01$\\
    Dec. 7  & 1.973 & $3.4\pm0.3$   & $0.86\pm0.05$ & $1.63\pm0.09$ & $0.27\pm0.02$ & ---\\
    Dec. 17 & 2.246 & $1.2\pm0.1$   & $0.31\pm0.02$ & $0.62\pm0.03$ & $0.08\pm0.01$ & ---          \\
    Dec. 19 & 2.306 & $1.6\pm0.2$   & $0.44\pm0.03$ & $0.78\pm0.05$ & $0.09\pm0.02$ & ---          \\
    Dec. 20 & 2.333 & $0.83\pm0.06$ & $0.34\pm0.02$ & $0.39\pm0.02$ & ---           & ---          \\
    Dec. 23 & 2.423 & $0.65\pm0.05$ & $0.29\pm0.02$ & $0.44\pm0.02$ & ---           & ---          \\
    Dec. 26 & 2.509 & $0.53\pm0.04$ & ---           & ---           & ---           & ---          \\
    Dec. 28 & 2.570 & $0.48\pm0.05$ & ---           & ---           & ---           & ---          \\
    Jan. 11 & 3.002 & $0.18\pm0.02$ & ---           & ---           & ---           & ---          \\
    Jan. 20 & 3.293 & $0.08\pm0.01$ & ---           & ---           & ---           & ---          \\
    \enddata
\end{deluxetable*}

\begin{deluxetable*}{lcccccc}
    \tablenum{B2}
    \tablecaption{Parameters for Haser Modeling.\label{tab:para}}
    \tablewidth{0pt}
    \tablehead{
        \colhead{Species} & \colhead{Band} & \colhead{Bandpass} & \colhead{$\log g$-factor} & \colhead{$v_{\rm exp}$} & \colhead{$l_0$} & $l_1$ \\
        \colhead{} & \colhead{} & \colhead{(${\rm\AA}$)} & \colhead{(${\rm erg\,mol^{-1}\,s^{-1}}$)} & \colhead{(km\,s$^{-1}$)} & \colhead{(10$^4$\,km)} & \colhead{(10$^4$\,km)}
    }
    \startdata
    CN          & $\Delta\nu=0$ & 3840--3900 & $f(\dot{r}_{\rm h})$ & 0.85     & 1.7  & 30 \\
    C$_3$       & $\cdots$      & 3960--4140 & $-12.42$             & $\cdots$ & 0.31 & 15 \\
    C$_2$       & $\cdots$      & 4960--5200 & $-12.35$             & $\cdots$ & 2.5  & 12 \\
    CH          & $\cdots$      & 4250--4330 & $-12.98$             & $\cdots$ & 7.8  & 0.48 \\
    % \hline
    % Fe\,{\sc i} & ---           & ---        & fluor. model         & 0.85     & ---  & --- \\
    % Ni\,{\sc i} & ---           & ---        & $\cdots$             & $\cdots$ & ---  & --- \\
    \enddata
    \tablecomments{The $g$-factors, expansion velocity ($v_{\rm exp}$), and scale lengths ($l_0$, $l_1$) are adopted from \citet{Cochran:2012} and references therein. All parameters are quoted at 1\,au. The $g$-factors scale as $r_{\rm h}^{-2}$, except for CN, for which the Swings-effect is taken into account \citep{Tatum:1977}. The expansion velocity scales as $r_{\rm h}^{-0.5}$. The scale lengths scale as $r_{\rm h}^{2}$, except for $l_0$ of C$_2$, which scales as $r_{\rm h}^{2.5}$ \citep{Cochran:2012}.
    }
\end{deluxetable*}

\begin{deluxetable*}{lccc}
    \tablenum{B3}
    \tablecaption{Parameters for [O\,{\sc i}] $\lambda6300$ Modeling.\label{tab:OI}}
    \tablewidth{0pt}
    \tablehead{
        \colhead{Species} & \colhead{$\alpha$} & \colhead{$\tau$} & \colhead{$v_{\rm exp}$} \\
        \colhead{} & \colhead{} & \colhead{(10$^4$\,s)} & \colhead{(km\,s$^{-1}$)}
    }
    \startdata
    H$_2$O & 0.05  & 8.3 & 0.85\\
    CO$_2$ & 0.37  & 40  & 0.80\\
    CO     & 0.047 & 33  & 0.43\\
    \enddata
    \tablecomments{The photodissociative yields ($\alpha$) and lifetimes ($\tau$) are adopted from \citet{McKay:2012}. The expansion velocities ($v_{\rm exp}$) are taken from \citet{Tan:2026}, \citet{Belyakov:2026}, and \citet{Biver:2026} for H$_2$O, CO$_2$, and CO, respectively, to remain consistent with the adopted production rates (see \S\ref{subsec:OI}). All $\tau$ and $v_{\rm exp}$ values are quoted at 1\,au. The $\tau$ scale as $r_{\rm h}^{2}$, and the $v_{\rm exp}$ scale as $r_{\rm h}^{-0.5}$, except for H$_2$O, for which a constant velocity is adopted following \citet{Tan:2026}.}
\end{deluxetable*}

\clearpage

\section{Contamination analysis of the [O\,I] $\lambda$6300 emission} \label{app:OI}

\begin{figure*}[ht!]
    \centering
    \includegraphics[width=0.5\textwidth]{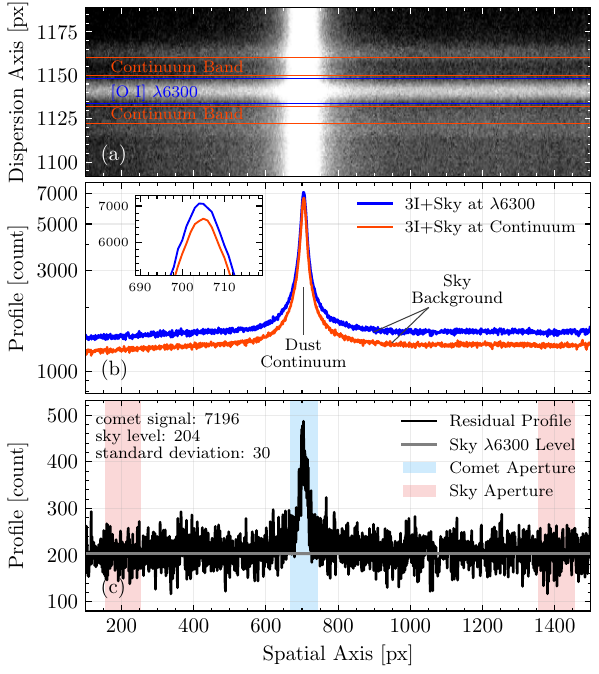}
    \caption{Illustration of the spatial-profile analysis around [O\,{\sc i}] $\lambda6300$ for the December 6 observation. (a) Two-dimensional spectrum of 3I near $\lambda6300$. The blue window is centered on $\lambda6300$, while the two red windows placed on nearby continuum regions. (b) Spatial profiles extracted from the $\lambda6300$ (blue) and continuum (red) windows. Both profiles contain the sky continuum and dust-scattered cometary continuum, whereas the $\lambda6300$ profile also contains sky and cometary $\lambda6300$ emission. The inset zooms in on the peak region. (c) The residual $\lambda6300$ profile after subtraction of the continuum profile. The sky emission contributes to the flat baseline, whereas the cometary emission produces the central excess. The gray line marks the sky $\lambda6300$ level, estimated as the mean value in the two sky apertures (red shaded regions) on either side of the comet, each 100 pixels wide. The blue shaded region marks the YFOSC extraction aperture, 78 pixels wide. \label{fig:profile}}
\end{figure*}

Although the $\lambda6300$ component has the higher SNR, it lies close to a telluric absorption feature that cannot be resolved at our spectral resolution, and may therefore be subject to contamination. In this case, the $I_{6300}/I_{6364}$ line ratio provides a useful consistency check. For our data, co-adding the first three spectra yields $I_{6300}/I_{6364}=3.0\pm0.3$, consistent with the theoretical optically thin value of 3.096 \citep{Galavis:1997}, indicating that telluric absorption is minor. 

Another concern is contamination from sky emission, especially when the sky $\lambda6300$ line is strong. To show this effect, we construct a residual spatial profile that contains only the sky and cometary $\lambda6300$ emission from the two-dimensional spectrum of 3I obtained on December 6. It is derived by taking the difference between the spatial profile of the $\lambda6300$ line and that of the nearby continuum region. The procedure is illustrated in Figure~\ref{fig:profile}, with the final residual profile shown in Figure~\ref{fig:profile}c.

% extract two spatial profiles from the two-dimensional spectrum of 3I obtained on December 6, with the adopted windows illustrated in Figure~\ref{fig:profile}a and the resulting profiles shown in Figure~\ref{fig:profile}b. Compared to the continuum profile that contains only the sky continuum and dust-scattered cometary continuum, the $\lambda6300$ profile contains additional sky and cometary $\lambda6300$ emission. We therefore subtract one from the other to remove the common continuum component and yield a residual profile that contains only the $\lambda6300$ emission, as shown in Figure~\ref{fig:profile}c. 

The sky $\lambda6300$ emission appears as a flat baseline in Figure~\ref{fig:profile}c and can affect the central cometary signal in two ways. First, Poisson noise from the sky background can increase statistical uncertainty. Second, imperfect estimation of the sky level can introduce a systematic bias. The former corresponds to the standard deviation of sky pixels, i.e., $\sigma_{\rm pixel}\approx30$ counts per pixel, while the latter (assuming a flat baseline) corresponds to the standard error of the mean sky level, which, by definition, is $\sigma_{\rm pixel}$ divided by the square root of the number of sky pixels $N_{\rm sky}$. In our data reduction pipeline, the sky background at each wavelength is estimated from two sky apertures on either side of the comet, each 100 pixels wide (see Figure~\ref{fig:profile}c). This gives $N_{\rm sky}=200$ pixels. 
The standard error is therefore $\sigma_{\rm mean}=\sigma_{\rm pixel}/\sqrt{N_{\rm sky}}\approx2$ counts per pixel. To evaluate their importance, we integrate the central cometary signal over the YFOSC extraction aperture, $N_{\rm comet}=78$ pixels, after baseline subtraction, which yields a total cometary signal of $\sim7200$ counts. The two effects therefore correspond to an uncertainty of $\sigma_{\rm pixel}\times\sqrt{N_{\rm comet}}\approx260$ counts and a possible bias of $\sigma_{\rm mean}\times N_{\rm comet}=156$ counts, or $\sim4$\% and $\sim2$\% of the integrated signal, respectively. We note that the former is already included in the quoted 7--12\% uncertainties (see \S\ref{subsec:OI}) through error propagation in our data reduction pipeline, while the latter is sufficiently small that it does not affect the results in \S\ref{subsec:OI}.

The third possible contaminant near the $\lambda6300$ line is the NH$_2$ (0,8,0) band. To evaluate this, we use the cleaner NH$_2$ (0,10,0) band, which is expected to have a comparable intensity \citep{Tegler:1989} but is less blended in our spectra. Integration over the NH$_2$ (0,10,0) band \citep[5675--5775\,${\rm\AA}$;][]{Langland-Shula:2011} gives no significant detection, with the measured intensity well below the 3$\sigma$ uncertainty. We therfore take the 3$\sigma$ uncertainty as an upper limit. After rescaling from the broader bandpass of NH$_2$ (0,8,0) band to the narrower wavelength interval around $\lambda6300$, we infer an NH$_2$ contribution of $<6$\%.

%% For this sample we use BibTeX plus aasjournalv7.bst to generate the

\bibliography{sample701}{}
\bibliographystyle{aasjournalv7}


\end{CJK*}
\end{document}